\input epsf

\newcount\counthoriz
\newcount\countvert


\newdimen\hstitle \newdimen\hsbody \newdimen\hsbox
\newdimen\hsgraph \newdimen\vsgraph
\tolerance=1000\hfuzz=2pt

\catcode`\@=11 

\hsbox=\hsize \advance\hsbox by-0.1truein
\hsgraph=\hsize \advance\hsgraph by-1.2truein
\vsgraph=\vsize \advance\vsgraph by-0.5truein

\def\nolabels{\def\wrlabeL##1{}\def\eqlabeL##1{}\def\reflabeL##1{}}
\def\writelabels{\def\wrlabeL##1{\leavevmode\vadjust{\rlap{\smash%
{\line{{\escapechar=` \hfill\rlap{\sevenrm\hskip.03in\string##1}}}}}}}%
\def\eqlabeL##1{{\escapechar-1\rlap{\sevenrm\hskip.05in\string##1}}}%
\def\reflabeL##1{\noexpand\llap{\noexpand\sevenrm\string\string\string##1}}}
\nolabels
 
\global\newcount\secno \global\secno=0
\global\newcount\meqno \global\meqno=1

\def\newsec#1{\bigskip\global\advance\secno
by1\message{(\the\secno. #1)}
\global\subsecno=0\eqnres@t
\global\figno=1
\global\tableno=1
\def\chap{#1}
\ifnum\lastpenalty>9000\else\bigbreak\fi
\noindent{{\bf\the\secno. #1}}
\par\nobreak\bigskip\nobreak}
\def\eqnres@t{\xdef\secsym{\the\secno.}\global\meqno=1\bigbreak\bigskip}
\def\sequentialequations{\def\eqnres@t{\bigbreak}}\xdef\secsym{}

\global\newcount\subsecno \global\subsecno=0
\def\subsec#1{\bigskip\global\advance\subsecno by1\message{(\secsym\the\subsecno. #1)}
\ifnum\lastpenalty>9000\else\bigbreak\fi
\noindent{{\bf\secsym\the\subsecno. #1}}\par\nobreak\bigskip\nobreak}

\def\eqn#1#2{\xdef #1{(\secsym\the\meqno)}\writedef{#1\leftbracket#1}%
\global\advance\meqno by1$$#2\eqno#1\eqlabeL#1$$}
\def\eqnmulti#1#2#3{\xdef #1{(\secsym\the\meqno)}\writedef{#1\leftbracket#1}%
\global\advance\meqno by1$$\eqalign{#2\cr #3}\eqno#1\eqlabeL#1$$}

\global\newcount\refno \global\refno=1
\newwrite\rfile

\def\ref{[\the\refno]\nref}
\def\nref#1{\xdef#1{[\the\refno]}\writedef{#1\leftbracket#1}%
\ifnum\refno=1\immediate\openout\rfile=refs.tmp\fi
\global\advance\refno by1\chardef\wfile=\rfile\immediate
\write\rfile{\noexpand\item{\bf #1\
}\reflabeL{#1\hskip.31in}\pctsign}\findarg}
\def\findarg#1#{\begingroup\obeylines\newlinechar=`\^^M\pass@rg}
{\obeylines\gdef\pass@rg#1{\writ@line\relax #1.^^M\hbox{}^^M}%
\gdef\writ@line#1^^M{\expandafter\toks0\expandafter{\striprel@x #1}%
\edef\next{\the\toks0}\ifx\next\em@rk\let\next=\endgroup\else\ifx\next\empty%
\else\immediate\write\wfile{\the\toks0}\fi\let\next=\writ@line\fi\next\relax}}
\def\striprel@x#1{} \def\em@rk{\hbox{}}

\newwrite\ffile\global\newcount\figno \global\figno=1
\newwrite\ffile\global\newcount\tableno \global\tableno=1

\newwrite\lfile
{\escapechar-1\xdef\pctsign{\string\%}\xdef\leftbracket{\string\{}
}

\def\writestoppt{}\def\writedef#1{}

\catcode`\@=12 

\def\del{\partial}
 \def\Tr{{\rm Tr}}
 \def\bar{\overline} 
\def\boxit#1{\vbox{\hrule\hbox{\vrule\kern3pt\vbox{\kern3pt#1\kern3pt}\kern3pt\vrule}\hrule}}

 \def\CS{{\cal S}} \def\CT{{\cal T}}
\def\vx{{\bf x}}

\def\vz{{\hbox{\font\tmp=cmsl10 \tmp z}}}

\def\frac#1#2{{{#1}\over {#2}}}
\def\tfrac#1#2{{\textstyle{{#1}\over {#2}}}}
\def\third{\hbox{${1\over 3}$}}

\def\PR#1#2#3{{\it Phys.~Rev.~}{\bf #1} (#2) #3}
\def\PRL#1#2#3{{\it Phys.~Rev.~Lett.~}{\bf #1} (#2) #3}
\def\NP#1#2#3{{\it Nucl.~Phys.~}{\bf #1} (#2) #3}
\def\PL#1#2#3{{\it Phys.~Lett.~}{\bf #1} (#2) #3}

\def\AP#1#2#3{{\it Ann.~Phys.~}{\bf #1} (#2) #3}

\def\figin{\epsfcheck\figin}\def\figins{\epsfcheck\figins}
\def\epsfcheck{\ifx\epsfbox\UnDeFiNeD
\message{(NO epsf.tex, FIGURES WILL BE IGNORED)}
\gdef\figin##1{\vskip2in}\gdef\figins##1{\hskip.5in}
\else\message{(FIGURES WILL BE INCLUDED)}
\gdef\figin##1{##1}\gdef\figins##1{##1}\fi}
\def\DefWarn#1{}
\def\figinsert{\goodbreak\midinsert}

\def\ifig#1#2#3{
\DefWarn#1\xdef#1{fig.~\secsym\the\figno}
\writedef{#1\leftbracket fig.\noexpand~\secsym\the\figno}
\figinsert\figin{\centerline{#3}}
\medskip\centerline{\vbox{\baselineskip12pt
\advance\hsize by -1truein\noindent{\bf
Fig.~\secsym\the\figno:} #2}}
\par\nobreak\bigskip\endinsert\global\advance\figno
by1}
\def\ifigp#1#2#3{
\DefWarn#1\xdef#1{fig.~\secsym\the\figno}
\writedef{#1\leftbracket fig.\noexpand~\secsym\the\figno}
\pageinsert\figin{\centerline{#3}}
\medskip\centerline{\vbox{\baselineskip12pt
\advance\hsize by -1truein\noindent{\bf
Fig.~\secsym\the\figno:} #2}}
\par\nobreak\bigskip\endinsert\global\advance\figno
by1}

\hyphenpenalty=10000
\font\tenmsbm=msbm10

\font\bigrm=cmr10 scaled\magstep2
\font\bigit=cmti10 scaled\magstep2
\font\bigbf=cmbx10 scaled\magstep2

\hfill\hbox{OUTP-96-21P}
\vskip 0.5in
\centerline{\bigbf High-Temperature Properties of the Z(3) Interface}
\centerline{\bigbf in (2+1)-D SU(3) Gauge Theory}
\vskip 0.5in
\centerline{\bigbf S.T. West and
J.F. Wheater\footnote*{\rm Email: west@thphys.ox.ac.uk,
jfw@thphys.ox.ac.uk}}
\vskip 0.5in
\centerline{\bigit Theoretical Physics,}\vskip3pt
\centerline{\bigit University of Oxford,}\vskip3pt
\centerline{\bigit 1, Keble Road,}\vskip3pt
\centerline{\bigit Oxford OX1 3NP,}\vskip3pt
\centerline{\bigit England, U.K.}


\vskip 2in
\centerline{\bigbf ABSTRACT}
\vskip 0.3in

\centerline{\vbox{\advance\hsize by -0.8truein\advance\baselineskip by
4pt{\noindent\bigrm
We study the high-temperature properties of the {\bigit Z(3)}
interface which forms between the various ordered phases of pure
{\bigit SU(3)} gauge theory above a critical temperature. On a (2+1)-D
Euclidean lattice, we perform an accurate measurement of the interface
tension, which shows good agreement with the prediction of
perturbation theory. We also examine the behaviour of the Debye
electric screening mass, and compare this with theoretical predictions.}}}

\vfill\supereject\vbox to \vsize{}

It has been known for some time that pure gauge theories have a
non-confining high-temperature phase \ref\IRoo{A.~Polyakov,
\PL{B72}{1978}{477}}. Therefore, any theory in a
confining phase at zero temperature ($T=0$) must have a phase transition at
some finite critical temperature, $T_c$, separating
the confining and non-confining phases.
A consideration of QCD at finite temperatures reveals that
the pure gauge $SU(3)$ sector has the two expected
phases. At low temperatures, there is the ``disordered'' phase, where the
colour charge of QCD is confined and the vacuum is symmetric under
the group $Z(3)$, the centre of the $SU(3)$ group. This is why no
free quarks are seen in the relatively cool universe of today.
At high temperatures, there is a
non-confining, $Z(3)$-breaking ``ordered'' phase, corresponding to
the free quark-gluon plasma believed to exist in the hot early
universe. Owing to the $Z(3)$ breaking, there exist three different,
degenerate high-temperature phases,
corresponding to the three members of $Z(3)$, the cube roots of
unity. Fermionic matter breaks the vacuum degeneracy, albeit on a
small scale, so we consider only the pure gauge theory in what follows.

Two different types of interface are possible
in the theory. First, there is one between the ordered and
disordered phases; this type is only stable at the critical
temperature, since only then do the temperature and pressure match
on both of its sides. Second, an interface can form between two of the
ordered phases with different $Z(3)$ vacua, and this is usually called
an ``order-order interface'', or ``$Z(3)$ interface''. All
thermodynamic quantities are the same on both sides of a $Z(3)$
interface, but they dip slightly across the object itself. Clearly,
this second type of interface can only exist above the critical
temperature.

It has recently been emphasised \ref\IRoba{V.~M.~Belyaev,
I.~I.~Kogan, G.~W.~Semenoff and N.~Weiss,
\PL{B277}{1992}{331}}\ref\IRoc{N.~Weiss, UBCTP93-23
(hep-ph/9311233)}\ref\IRod{I.~I.~Kogan, \PR{D49}{1994}{6799}} that the
order parameter
distinguishing the $Z(3)$ phases is fundamentally a Euclidean object,
with no counterpart in Minkowski space. Thus, whilst the $Z(3)$
symmetry exists, and hence $Z(3)$ domains and interfaces exist, in the
Euclidean path integral, it is less certain that the interfaces can exist
as physical objects in the universe. The claim has even been
made \ref\IRoe{A.~V.~Smilga, \AP{234}{1994}{1}} that only one true physical
phase exists, even in Euclidean space, at high temperatures, and we
shall address this when discussing our results later. In Minkowski
space, the interfaces appear \IRoc\ to
have unphysical thermodynamic properties when a certain number of
fermion families is considered in addition to the gauge fields,
resulting from the unusual fact that the ordered phases occur at {\it high}
temperatures.
Even setting this problem aside, the problem of interpretation of the
interface remains, as the order parameter is non-local in Euclidean
time, corresponding to an imaginary time-like gauge field in Minkowski
space, and thus an imaginary chemical potential for the colour
charge \ref\IRof{A.~Roberge and N.~Weiss, \NP{B275}{1986}{734}}. However,
by virtue of their contribution to the partition function
and to expectation values calculated using the Euclidean path
integral, the interfaces must be included (like the instanton) in a
non-perturbative analysis of the thermodynamics of QCD.

In the Euclidean formalism, the ``time'' dimension runs from $0$ to
$\beta_T=1/T$. The phases of a thermal gauge theory are characterised
by the free energies of static configurations of quarks and
antiquarks \ref\IRv{L.~D.~McLerran and B.~Svetitsky, \PR{D24}{1981}{450}}.
In particular, the self-energy of a single quark in the gluon medium, $F_q$,
is given by the expectation value of a single Polyakov line, 
$L(\vx)$, a time-ordered Wilson loop which wraps
around the time boundary for a fixed spatial location $\vx$:
\eqn\IExiv
{L(\vx)=\third\Tr\CT e^{ig\int_0^{1/T} d\tau
A_0(\tau,\vx)}=e^{-{\beta_T} F_q}.}
If $<L(\vx)>=0$ then the insertion of a single quark will require
infinite energy, corresponding to a confining phase. In contrast, if
$<L(\vx)>\neq 0$ then the insertion energy will be finite and the colour
charge deconfined. Thus, the Polyakov line gives an effective test for
confinement; unlike topologically trivial Wilson loops,
it is aware of the phase structure of an $SU(3)$
system and can map out the structure of any interface.

A general non-Abelian gauge transformation which leaves the Euclidean
QCD action, $\CS_E$, invariant will also leave $L$ invariant if the
transformation is periodic in time. However, 
$\CS_E$ is actually invariant under a larger
group than merely the periodic gauge transformations, being unaffected
by transformations belonging to the 
centre of the gauge group (the set of
elements which commute with all members of the group, $Z(N)$ in the
case of $SU(N)$). This global symmetry
is an invariance under gauge transformations which are only periodic
up to an element of the centre; it leaves local observables unchanged, and
it remains after all gauge fixing. However, the
topologically non-trivial
Polyakov line, which wraps around the periodic boundary condition in
the time direction, is not invariant under these transformations, but rather
is rotated by an element, \vz, of the centre: $L(\vx) \rightarrow \vz L(\vx).$
Clearly, $<L(\vx)>$ acts as an order parameter for the centre
symmetry, distinguishing between broken and unbroken phases, and
between the different broken phases. This has been seen
perturbatively \ref\IRviioa{N.~Weiss,
\PR{D24}{1981}{475}}\ref\IRx{T.~Bhattacharya, A.~Gocksch, C.~P.~Korthals~Altes
and R.~D.~Pisarski, \NP{B383}{1992}{497}} and by simulation \IRv.

In the deconfined phase, where the $Z(N)$ symmetry has been spontaneously
broken, the system can exist in any one of $N$
degenerate vacua. It is possible to arrange the boundary conditions
of the system so that different parts of it exist in different
vacua, {\it i.e.} different $Z(N)$ phases. The existence of these
distinct domains forces the appearance of domain walls, or ``$Z(N)$
interfaces'', where they meet \ref\IRix{T.~Bhattacharya, A.~Gocksch,
C.~P.~Korthals~Altes and R.~D.~Pisarski, \PRL{66}{1991}{998}}\IRx.
Within these interfaces, the gauge
fields interpolate between the different vacua, as does the
expectation value of the Polyakov line.

The calculation of the interface tension is an instanton
problem, in which $L(\vx)$ interpolates from one $Z(3)$ vacuum to
another. An effective one-dimensional theory will describe the
interface profile. The instanton is the solution of the classical
equations of motion, with action $\sim 1/g^2$ (as seen by absorbing
the dimensionless coupling through $gA^\mu\rightarrow A^\mu$), and so we
might na\"{\i}vely expect the interface tension to have the same
behaviour. To check this, 
an effective action can be constucted from the classical piece, acting as a
kinetic term, and a quantum piece, a potential term formed by
integrating out fluctuations to one-loop order. The $Z(N)$ instanton is
the stationary point of the total action.
The instanton calculation has been performed
\IRix\IRx\ref\IIRx{C.~P.~Korthals~Altes, A.~Michels, M.~Stephanov and
M.~Teper, \NP{B (Proc. Suppl.) 42}{1995}{517}}, predicting that,
in 2+1 Euclidean dimensions, the interface tension (the effective action
per unit area of the interface) takes the form $\alpha=\alpha_0 T^{2.5}/g$,
where $\alpha_0$ is a constant predicted to be 8.33 for $SU(3)$ in the
continuum, in the limit as $T\rightarrow\infty$.

In the lattice formalism, one considers a hypercubic lattice whose sites are
separated by lattice spacing $a$. Let
us suppose there are two space dimensions,
one ($x$) much longer than the other ($y$), and both much longer than
the time dimension: $L_x \gg L_y \gg {\beta_T}$.
If we specify $N_t$ links
in the time direction, this corresponds to a temperature
$T=1/\beta_T=1/N_t a$; similarly, $L_y=N_y a$ and $L_x=N_x a$.
The conventional bare (dimensionless) coupling, $g$,
is related to the lattice parameter, $\beta$, by
${\beta = 2N/a^{4-d}g^2}$
in $d$ dimensions. If $d=3$ and $N=3$ then 
$\beta=6/ag^2$, and for $N_t=2$ a lattice calculation
\ref\IRxiia{C.~P.~Korthals~Altes,
A.~Michels, M.~Stephanov and M.~Teper, OUTP-96-10P} predicts that
\eqn\IExxi
{\alpha_0=9.82.}
To test these predictions, we have performed extensive simulations of
the (2+1)-D $SU(3)$ theory, where the interface is one-dimensional (a string).
A similar prediction for (2+1)-D $SU(2)$ gauge theory was borne out
for a range of lattice sizes with $N_t\leq 5$ in a recent
survey \IIRx.

We impose periodic boundary conditions in
all directions, and set the temperature, specified through the lattice
parameter, to be $\beta \gg \beta_c$, where $\beta_c$ corresponds to
the temperature of the deconfinement phase transition, previously
estimated \ref\IRxv{J.~Christensen,
G.~Thorleifsson, P.~H.~Damgaard and J.~F.~Wheater, \PL{276B}{1992}{472}}
as $\beta_c=8.175(2)$.
Left to its own devices, the entire system will settle into just one of the
$Z(3)$ phases, this being the configuration of lowest energy.
We need to force the
system into producing a $Z(3)$ interface, and this is achieved by
use of a ``twist'' \ref\IIRo{K.~Kajantie, L.~K\"arkk\"ainen
and K.~Rummukainen, \NP{B357}{1991}{693}}. 

The twist is introduced by modifying the action so that
every plaquette in the $\tau-x$ plane at some particular value of
$\tau$ and $x$ is to be premultiplied by an element, $\vz^{-1}$, of
the centre group when it appears in the action. This means that if we
order the 
system so that all the Polyakov lines to the left of the twist
are in the ${\hbox{\tenmsbm Z}}(3)$ phase given by $<L>=1$,
then the twist 
will cause the Polyakov lines to take the value $<L>=\vz$
to the right of the twist. No extra energy is associated with the
twist itself; it merely acts as a change of variables,
transforming one 
$Z(3)$ vacuum into another. However, the periodic boundary conditions
force the appearance of a real physical $Z(3)$ interface somewhere else in
the $x$ direction, where the Polyakov lines interpolate from $\vz$
back to $1$. Thus, we have forced the creation of two $Z(3)$ domains
on the lattice, with one $Z(3)$ interface between them. We
place the twist at the boundary in our simulations, so that the
low-$x$ end of the system is in one phase and the high-$x$ end in another.

For our Monte-Carlo simulations, we used a mix of Cabibbo-Marinari heat-bath
steps \ref\IIRiv{N.~Cabibbo and E.~Marinari,
\PL{119B}{1982}{387}}, using the Kennedy-Pendleton
algorithm \ref\IIRii{A.~D.~Kennedy and B.~J.~Pendleton,
\PL{156B}{1985}{393}} to update $SU(2)$
subgroups, and Brown-Woch over-relaxation steps \ref\IIRvi{F.~R.~Brown
and T.~J.~Woch, \PRL{58}{1987}{2394}}, after using
many initial heat-bath sweeps to equilibrate the system.
Measurements of
physical operators took place only every four sweeps, to reduce the
correlations caused by adjacent configurations. All calculations were
performed on a DEC 2100 A500MP machine, with CPU times tabulated later
in table 1.

\ifig\IIFia{The real and imaginary parts of Polyakov lines
on a $2\times 24\times 96$ lattice with twist.}
{\epsfxsize\hsgraph \epsfbox{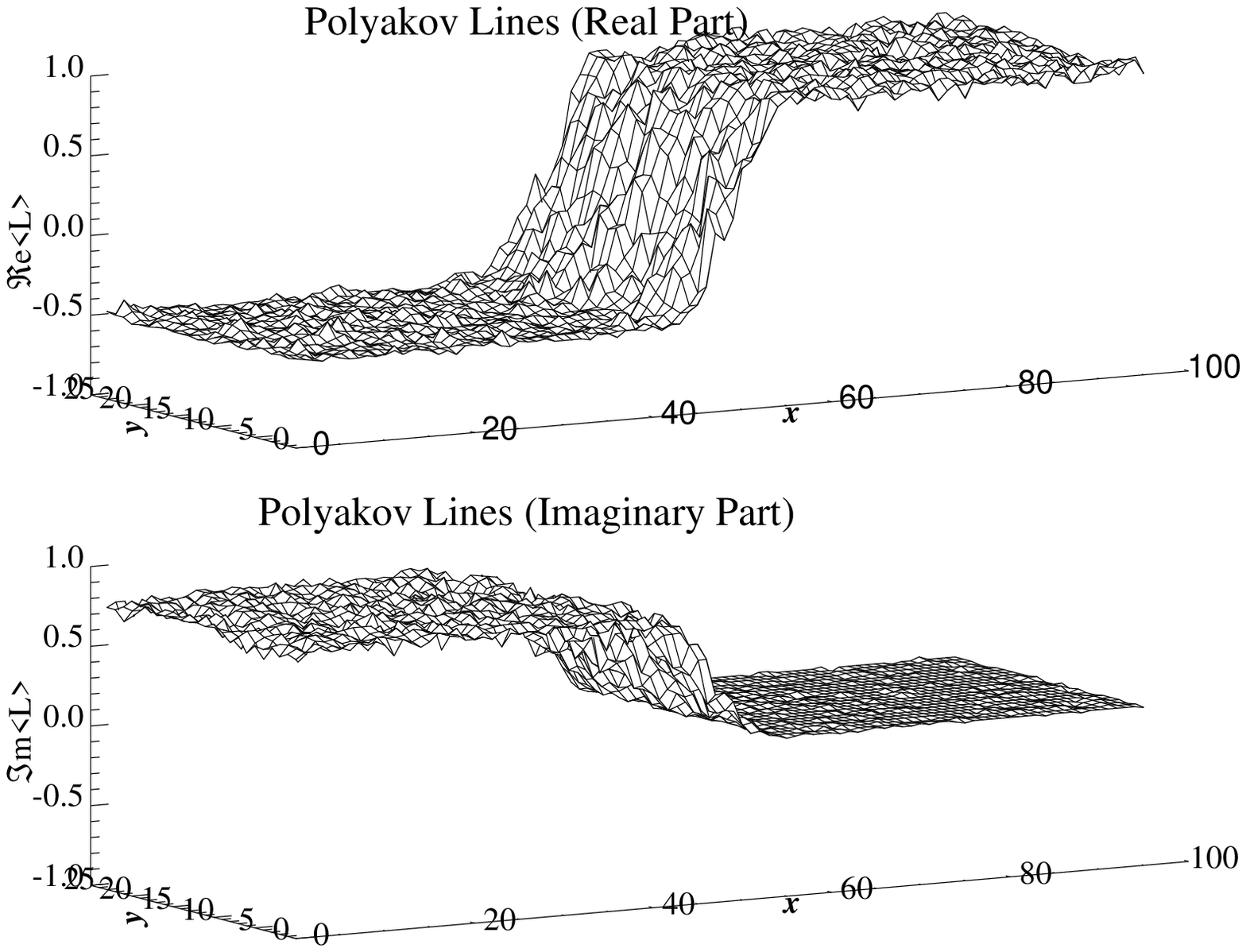}}

In \IIFia, an interface is shown on a $2\times 24 \times 96$ lattice.
The longitudinal direction goes across
the page, and the transverse direction into it. The ``height'' in each
case is determined by the value of the Polyakov line, and has nothing
to do with the time dimension, which has been integrated over to form
the Polyakov line.
The twist has been placed at the
boundary of the $x$ direction, as mentioned before, so the left-hand
end of the lattice is in one $Z(3)$ phase with $<L>\approx e^{2\pi
i/3}$, and the right-hand end in another with $<L>\approx 1$. The
temperature is fairly high: $\beta =50 \gg \beta_c$. Thus, the
interface is fairly well-defined and stable, since its free energy
is correspondingly high which suppresses fluctuations in
phase.
In this picture, the
interface is roughly in the centre of the lattice. However, the
interface is actually translation invariant; there is nothing in the
action to specify that it should be located at any particular value of
$x$, and it can be seen to execute a random walk along the lattice,
with a speed which increases slightly as the temperature decreases.

The free energy of a system is $F=-T\ln Z$, so
$\tfrac{\del}{\del\beta}
(\tfrac{F}{T}) = \tfrac{1}{\beta} <S>.$
This gives the behaviour of the free energy in terms of the average
plaquette action. Simulations are performed for the same lattice with
and without twisted plaquettes, {\it i.e.} with and without a $Z(3)$
interface, and the difference in free energy
between the two gives the free energy of the interface itself, from
which we can measure its surface tension:
$F_{interface} = F_{twisted} - F_{untwisted} = \alpha A,$
where $\alpha$ is the surface tension, and $A$ is the `area' of the
interface (in our case, the transverse size of the lattice).
Therefore, substituting $A=L_y a$ into the previous equation,
\eqn\IIEiv
{\frac{\Delta S}{\beta}\quad=\quad\frac{\del}{\del\beta} \bigg (\frac{\alpha_0
T^{2.5}}{g} \frac{L_y a}{T} \bigg
)\quad=\quad\frac{L_y}{2\sqrt{6\beta}\ 
N_t^{1.5}}\ \alpha_0.}

The prediction \IExxi\ that $\alpha_0=9.82$ for a system with $N_t=2$
was made in the infinite-temperature limit. Hence, to test this prediction
it is necessary to estimate $\alpha_0$ in this limit. The best way to
do this is to carry out a survey of $\alpha_0$ for various values of
$\beta$, and then to extrapolate the results to the infinite-$\beta$ limit.
Several tens of thousands of heat-bath sweeps were used to ensure complete
equilibration of the lattice before measurements began. ``Main
sweep'' refers to a sweep after equilibration is complete.

\bigskip\boxit{
\vbox{\tabskip=0pt \offinterlineskip
\def\tabletitles#1#2#3#4#5#6{\DefWarn#1\xdef#1{table
\secsym\the\tableno}&&\multispan7\hfil {Table
\secsym\the\tableno: #2}\hfil&\cr \tableedge\tablerule\tablelines
&&#3&&#4&&#5&&#6&\cr}
\def\tablerule{\noalign{\hrule}}
\def\tablelines{\omit&height2pt&\omit&height2pt&\omit&height2pt&\omit&height2pt&\omit&height2pt\cr}
\def\tableedge{\omit&height2pt&\multispan7&\cr}
\def\tablegap{\omit&\omit\vbox to 3pt{}&\multispan7\cr}
\halign to\hsbox{\strut#& \vrule#\tabskip=1em plus 2em& \hfil#\hfil &
\vrule# & \hfil#\hfil & \vrule# & \hfil#\hfil & \vrule# & \hfil#\hfil &
\vrule#\tabskip=0pt\cr\tablerule\tableedge
\tabletitles{\IITi}{Simulation Details}{$\beta$}{Lattice Size}{Heat-Bath
+ Main Sweeps (000's)}{CPU Time (hrs)}
\tablelines\tablerule\tablegap\tablerule\tablelines
&&16&&$2\times 12\times 30$&&15 + 400&&80&\cr
\tablelines\tablerule\tablelines
&&28.125&&$2\times 12\times 36$&&20 + 400&&120&\cr
\tablelines\tablerule\tablelines
&&50&&$2\times 16\times 48$&&20 + 400&&260&\cr
\tablelines\tablerule\tablelines
&&112.5&&$2\times 24\times 72$&&40 + 400&&800&\cr
\tablelines\tablerule\tablelines
&&200&&$2\times 32\times 96$&&40 + 400&&1600&\cr
\tablelines\tablerule}}\global\advance\tableno by1}\bigskip

In table 2, we list the
results for the average action for a single plaquette,
$<S^{\hbox{\tenmsbm P}}>$, the total action divided by $3 N_t N_x N_y$.
The difference between the twisted and untwisted
values is then used in \IIEiv\ to obtain the
measurements of $\alpha_0$ in the last column.

\bigskip\boxit{
\vbox{\tabskip=0pt \offinterlineskip
\def\tabletitles#1#2#3#4#5#6{\DefWarn#1\xdef#1{table
\secsym\the\tableno}&&\multispan7\hfil {Table
\secsym\the\tableno: #2}\hfil&\cr \tableedge\tablerule\tablelines
&&#3&&#4&&#5&&#6&\cr}
\def\tablerule{\noalign{\hrule}}
\def\tablelines{\omit&height2pt&\omit&height2pt&\omit&height2pt&\omit&height2pt&\omit&height2pt\cr}
\def\tableedge{\omit&height2pt&\multispan7&\cr}
\def\tablegap{\omit&\omit\vbox to 3pt{}&\multispan7\cr}
\halign to\hsbox{\strut#& \vrule#\tabskip=1em plus 2em& \hfil#\hfil &
\vrule# & \hfil#\hfil & \vrule# & \hfil#\hfil & \vrule# & \hfil#\hfil &
\vrule#\tabskip=0pt\cr\tablerule\tableedge
\tabletitles{\IITii}{Simulation
Results}{$\beta$}{$<S^{\hbox{\tenmsbm P}}>_{twisted}$}{$<S^{\hbox{\tenmsbm P}}>_{untwisted}$}{$\alpha_0$}
\tablelines\tablerule\tablegap\tablerule\tablelines
&&16&&0.17510(1)&&0.17379(1)&&13.1(1)&\cr
\tablelines\tablerule\tablelines
&&28.125&&0.097575(6)&&0.096832(6)&&11.79(13)&\cr
\tablelines\tablerule\tablelines
&&50&&0.054310(3)&&0.053916(3)&&11.12(12)&\cr
\tablelines\tablerule\tablelines
&&112.5&&0.0239740(7)&&0.0238074(7)&&10.58(6)&\cr
\tablelines\tablerule\tablelines
&&200&&0.0134549(5)&&0.0133644(5)&&10.21(8)&\cr
\tablelines\tablerule}}\global\advance\tableno by1}\bigskip

One would not expect the finite size of the lattices used to have a
great effect on the untwisted results, as the whole lattice will be in
one phase, but it is less clear how great the effect would be in the
presence of the twist. The shape of the interface may not
flatten out completely at opposite ends of the longitudinal dimension if
the lattice is too short, skewing our measurement
of the total interface free energy. The lattice sizes above were
chosen to be large
enough for the finite-size effects to be negligible, by studying
results for $\beta=50$ on various sizes of lattice and scaling the
spatial dimensions with the interface width, which is proportional to
$\sqrt\beta$ and which we choose
as our measure of physical correlation lengths. The only exception to
this was the measurement for $\beta=16$, for which the
lattice could not safely be made any smaller.
The results are given in table 3, each
from 2+20 thousand sweeps. Similar runs without twist gave a 
result of $<S^{\hbox{\tenmsbm P}}>_{untwisted} \approx
0.05390(1)$, and this value was assumed for each lattice in the
calculation of $\alpha_0$.
The results suggest, given the errors, that a lattice size of $2\times
16\times 48$ is adequate for $\beta=50$.

\bigskip\boxit{
\vbox{\tabskip=0pt \offinterlineskip
\def\tabletitles#1#2#3#4#5{\DefWarn#1\xdef#1{table
\secsym\the\tableno}&&\multispan5\hfil {Table
\secsym\the\tableno: #2}\hfil&\cr \tableedge\tablerule\tablelines
&&#3&&#4&&#5&\cr}
\def\tablerule{\noalign{\hrule}}
\def\tablelines{\omit&height2pt&\omit&height2pt&\omit&height2pt&\omit&height2pt\cr}
\def\tableedge{\omit&height2pt&\multispan5&\cr}
\def\tablegap{\omit&\omit\vbox to 3pt{}&\multispan5\cr}
\halign to\hsbox{\strut#& \vrule#\tabskip=1em plus 2em& \hfil#\hfil &
\vrule# & \hfil#\hfil & \vrule# & \hfil#\hfil &
\vrule#\tabskip=0pt\cr\tablerule\tableedge
\tabletitles{\IITiii}{Finite-Size Survey: $\beta=50$}
{Lattice Size}{$<S^{\hbox{\tenmsbm P}}>_{twisted}$}{$\alpha_0$}
\tablelines\tablerule\tablegap\tablerule\tablelines
&&$2\times 16\times 32$&&0.05449(1)&&11.1(3)&\cr
\tablelines\tablerule\tablelines
&&$2\times 16\times 40$&&0.05438(1)&&11.3(3)&\cr
\tablelines\tablerule\tablelines
&&$2\times 16\times 48$&&0.05432(1)&&11.9(3)&\cr
\tablelines\tablerule\tablelines
&&$2\times 16\times 64$&&0.05422(1)&&12.0(3)&\cr
\tablelines\tablerule\tablelines
&&$2\times 24\times 48$&&0.05432(1)&&11.9(3)&\cr
\tablelines\tablerule\tablelines
&&$2\times 24\times 60$&&0.05423(1)&&11.6(3)&\cr
\tablelines\tablerule\tablelines
&&$2\times 24\times 72$&&0.05418(1)&&11.9(3)&\cr
\tablelines\tablerule}}\global\advance\tableno by1}\bigskip

Now, as mentioned above, the prediction of \IExxi\ is only valid in the limit
$\beta\rightarrow\infty$, so the question arises of how to extrapolate
our results to test this prediction. This is really a question of
calculating infra-red divergences.
In $3+1$ dimensions, one would
expect the one-loop correction to \IExxi\ to go like $1/\beta$ in
this limit, since the Debye mass, $m\sim T$. However, in
$2+1$ dimensions, it is conceivable that a dependence on $\ln\beta$
would need to be taken into account, as this appears in the infra-red
correction for the Debye mass, discussed later. For this reason,
it is difficult to predict the form of the $\beta$-dependence. In any
case, it would be
quite possible for the logarithmic dependence to masquerade as a power
law over a restricted range of $\beta$.

\ifig\IIFivo{Extrapolations of $\alpha_0$ are shown as
$\beta\rightarrow\infty$ for various powers of $\beta$, with each
extrapolation labelled accordingly:
the five data points are plotted as triangles for an x-axis of
$1/\beta^{0.95}$; as squares for the preferred power,
$1/\beta^{0.85}$; and as diamonds for $1/\beta^{0.75}$.}
{\epsfxsize\hsize \epsfbox{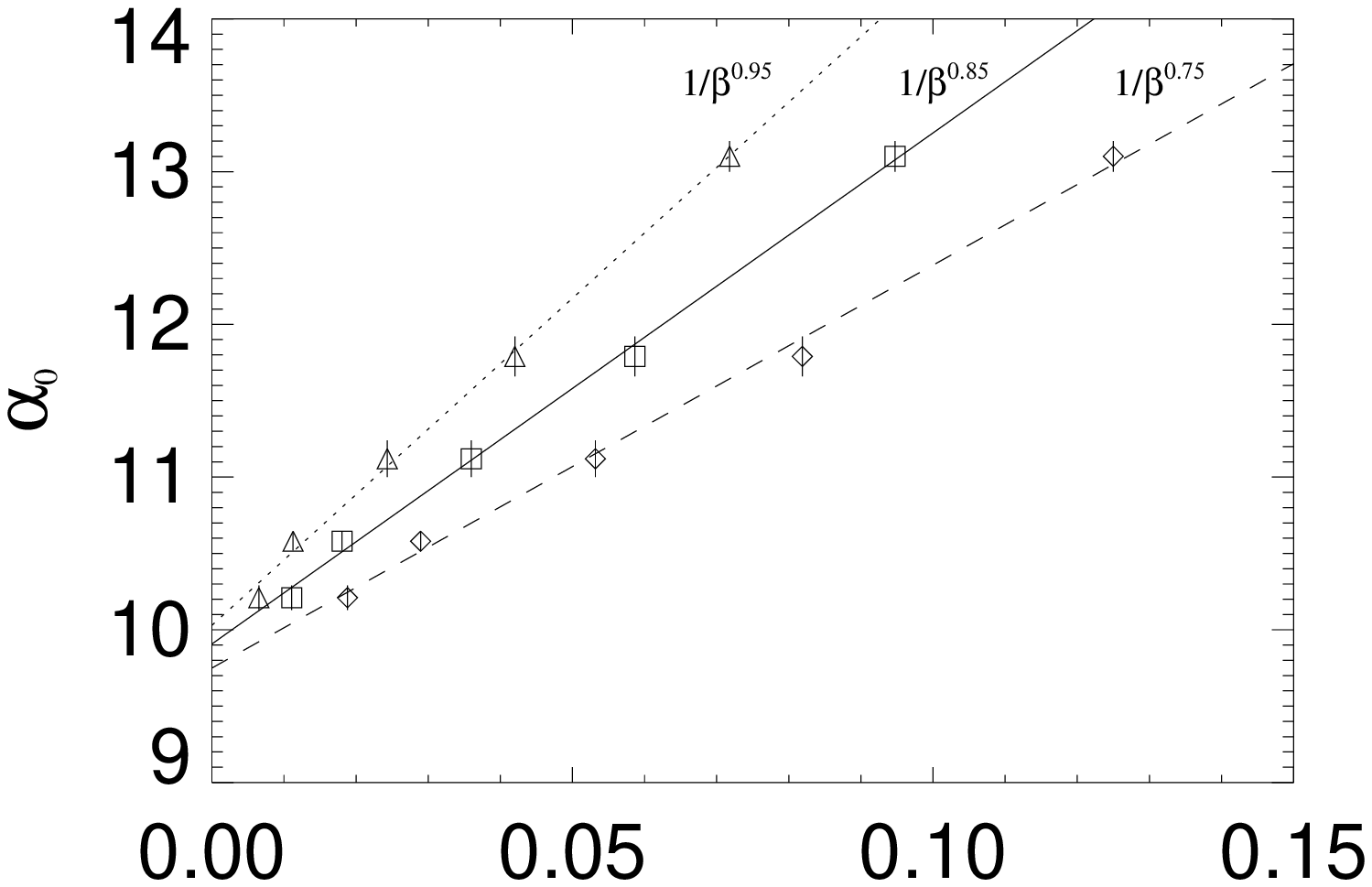}} 

In the absence of a definite prediction for the form of the
finite-$\beta$ correction, we perform the extrapolation using a best
fit to the form $\alpha_0(\beta)=\alpha_0(1+c/\beta^w)$.
A computer-generated fit to the data,
leaving the power, $w$, as a free parameter, suggests a
power of $0.85$ with $\chi^2=2.39$. This gives a $\beta=\infty$ limit of
$\alpha_0=9.91(1)$, as extrapolated in \IIFivo. Fits for various
imposed powers
give a lowest $\chi^2$ of $2.34$ for a power of $0.80$, with
$\alpha_0=9.83(1)$. However,
$\chi^2\leq 2.6$ for powers between $0.75$ and $0.90$. Therefore, in
the absence of a definite prediction for the power of $1/\beta$, we
allow for a reasonable error in the fit, and use $0.85(10)$, the range
for which extrapolations are illustrated in \IIFivo.
Thus, our infinite-temperature extrapolation for the interface tension
is
$$\alpha_0=9.91(1)(14),$$ in very good agreement with the instanton
calculation result of \IExxi. The first error is associated with the
fit for $1/\beta^{0.85}$, and the second with the uncertainty in that
value for the power.

It has been argued \IRoe\ 
that the different $Z(3)$ vacua distinguished by different values of
the Polyakov line actually correspond to one and the same physical
state. This claim comes from considering the r\^ole of infra-red
divergences in the calculation of the surface tension, and, if true,
would imply that the surface tension is actually zero (since no
physical interface can exist). The agreement of our results for
$\alpha_0$ with perturbation theory tends to disprove this
argument. It is, of course, conceivable that our measurements for
$\alpha_0$ would dramatically fall towards zero if we went to higher
and higher $\beta$, meaning that our extrapolation is simply an
artefact for the range of $\beta$ that we have considered. However,
our highest value is $\beta\approx 25\beta_c$, a temperature well
above the critical point, and it seems unlikely that any dramatically
different behaviour
will set in at a temperature higher than this. A sceptic could
also argue that a lattice with $N_t=2$ could show behaviour quite
distinct from that in the continuum; but quite apart from the agreement with
the lattice prediction of $\alpha_0$ for this value of $N_t$, a
similar survey carried out for $SU(2)$ gauge theory for a range of
$N_t\leq 5$ has found similar agreement with perturbation
theory \IIRx.

The electric (``Debye'') mass, or inverse Debye screening length, $m$,
governs gauge-invariant correlation functions of the time-like
component of the gluon field ($A_0$) at large distances and high
temperatures. The free energy of a quark-antiquark
pair, over and above the sum of their separate free energies,
vanishes as the quarks become infinitely far apart,
and, from perturbation theory, we know that quarks are screened at a
distance of the order of the inverse electric mass, so rather than
being logarithmic in $\vx$, the interaction potential takes
the form
$$F_{q\bar
q}(\vx)-2F_q=V(|\vx|,T)\mathrel{\mathop{\sim}\limits_{|\vx|\rightarrow
\infty}} -Ce^{-2m|\vx|} \qquad\hbox{for\ } T>T_c.$$
The factor of two arises because gauge invariance leads to an exchange of
two gluons being the lowest-order contribution in perturbation theory.
The electric mass is one of the fundamental parameters of a gauge
theory at finite temperature.
A self-consistent and gauge-invariant calculation of its
value \ref\IRxiii{E.~D'Hoker,
\NP{B200[FS4]}{1982}{517}}\ref\IRxiv{E.~D'Hoker, \NP{B201}{1982}{401}}
gives
\eqn\IExxii
{m^2=\frac{g^2NT}{4\pi}\ln(T/g^2)\quad+\quad
O\bigg[g^2T\ln(\ln(T/g^2))\bigg],\qquad T/g^2\gg 1,}
where the dimensional regularisation mass scale is set to $m$.
Clearly, this mass is a non-analytic function of coupling
constant and temperature.

To measure the Debye mass, we use the correlation of
the transverse average of Polyakov lines on the untwisted lattice.
We estimate $2m$ from the vacuum-subtracted
correlations by assuming an exponential decay with the line
separation, $x$, using the function
$$\mu(x)=\ln\bigg(\frac{<L^\dagger(x-1)L(0)>-<\bar
L>^2}{<L^\dagger(x)L(0)>-<\bar L>^2}\bigg),$$
where each $<L(x)>$ represents an average of $L(x)$ over all configurations
(sweeps), and $<\bar L>$ is a further average over all $x$. From the
definition of the Debye mass, we expect
$\mu(x)$ to tend to a constant, $2m$, as we increase the separation,
$x$, as long as $x$ remains significantly less than $L_x/2$.
For each set of data at a particular $\beta$, we extract a value of
$\mu(x)$ corresponding to the level at which the function flattens
out before errors begin to increase wildly. These values are given in
table 4 for the simulations listed in \IITi.

\bigskip\centerline{\boxit{
\vbox{\tabskip=0pt \offinterlineskip
\def\tabletitles#1#2#3#4#5#6#7#8#9{\DefWarn#1\xdef#1{table
\secsym\the\tableno}&&\multispan{13}\hfil {Table
\secsym\the\tableno: #2}\hfil&\cr \tablerule\tablegap\tablerule\tablelines
&&#3&&#4&&#5&&#6&&#7&&#8&&#9&\cr}
\def\tablerule{\noalign{\hrule}}
\def\tablelines{\omit&height2pt&\omit&height2pt&\omit&height2pt&\omit&height2pt&\omit&height2pt&\omit&height2pt&\omit&height2pt&\omit&height2pt\cr}
\def\tableedge{\omit&height2pt&\multispan{13}&\cr}
\def\tablegap{\omit&\omit\vbox to 3pt{}&\multispan{13}\cr}
\halign to\hsize{\strut#& \vrule#\tabskip=1em plus 2em& \hfil#\hfil &
\vrule#\tabskip=0pt&#& \vrule#\tabskip=1em plus 2em & \hfil#\hfil &
\vrule# & \hfil#\hfil & \vrule# & \hfil#\hfil & \vrule# & \hfil#\hfil &
\vrule# & \hfil#\hfil & \vrule#\tabskip=0pt\cr\tablerule\tableedge
\tabletitles{\IITiv}{Debye (Electric) Mass (units of $a^{-1}$)}
{$\beta$}{\hbox to 3pt{}}{16}{28.125}{50}{112.5}{200}
\tablelines\tablerule\tablelines
&&{Twice Debye mass, $\mu=2m$}&&&&{0.80(1)}&&{0.690(5)}&&{0.555(5)}&&{0.410(5)}&&{0.332(4)}&\cr
\tablelines\tablerule}}\global\advance\tableno by1}}\bigskip

\ifig\IIFv{Results for the Debye mass. The prediction of \IExxii\ is
plotted to leading-order only (dashes); with an
additional correction of $\ln(\ln(\beta/12))$ (dots); and with an
alternative additional correction of
$0.14\ln(\ln(\beta/12))+0.94$ (dot-dashes).}
{\epsfxsize\hsize \epsfbox{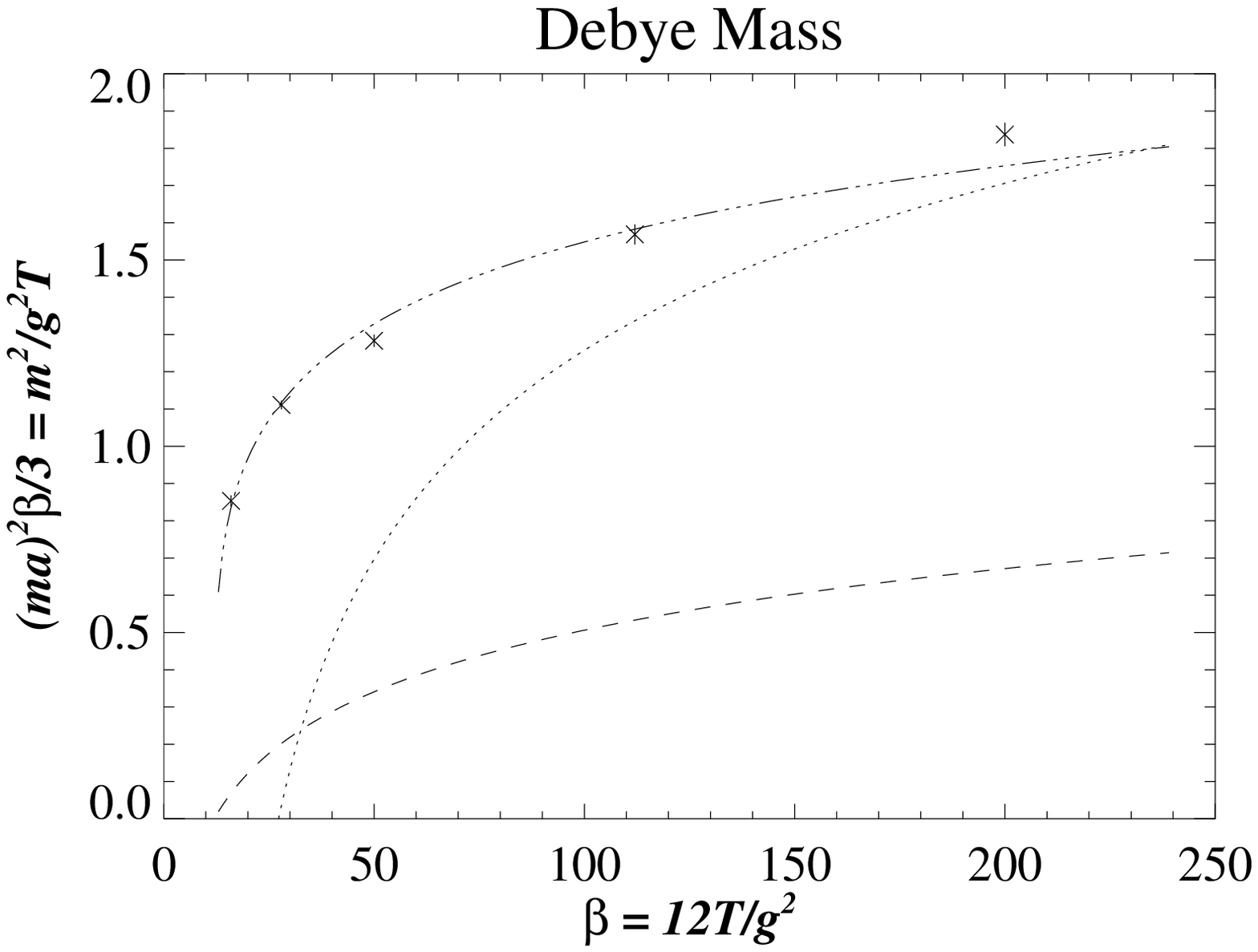}}

The values are plotted in \IIFv,
together with the theoretical prediction of \IExxii\ (dashed line).
The results show some qualitative agreement with the leading-order
prediction of \IExxii.
However, although the shape of the rise of the mass with
temperature looks correct, the results differ from the prediction by a
factor of two or so. The reason for this becomes clearer when we
consider the size of the $O[\ln(\ln(\beta/12))]$ correction
compared to the leading-order $\frac{3}{4\pi}\ln(\beta/12)$ that we
plot: for our range $16<\beta<200$, the leading-order term is between
$0.1$ and $0.7$, but the correction is $O(-1.2)$ to $O(1.0)$! Thus, it
is very difficult to draw any firm conclusions from the
data to support or deny the prediction.

To show the effect of the correction, the dotted curve in
\IIFv\ is $\frac{3}{4\pi}\ln(\beta/12)+\ln(\ln(\beta/12))$. Our
data certainly look more reasonable when this correction is included,
but, of course, this estimate is only of the order of the correction,
not of an exact expression. A free fit to a correction of the form
``$c_1\ln(\ln(\beta/12)) + c_2$'' gives the dot-dashed line in the
\IIFv, with $c_1\approx 0.14$ and $c_2\approx 0.94$, which shows
extremely good agreement with our data. Although we have two free
parameters and only five data points, the fit is strikingly good, and
this gives us some reason to hope that our data is consistent with the
theoretical prediction. Of course, no firm conclusions can be drawn
without a more explicit expression for the
$O[\ln(\ln(\beta/12))]$ correction. However, the ratio of
coefficients of the free fit does at least confirm that most of the
temperature dependence is contained within the leading
$\frac{3}{4\pi}\ln(\beta/12)$ term.

In conclusion, we have
measured the interface tension on the lattice at high temperatures,
and extrapolated our
results to the infinite-temperature limit, where we have found good
agreement with the prediction of an instanton calculation. This
reinforces the argument for the existence of $Z(3)$-breaking phases in
the Euclidean functional integral, although their physical existence
in the early universe remains in question, and suggests that the
predictions of perturbation theory are indeed reliable in this
regime. We have also examined the
Debye profile of the interface, but have not been able to compare our
results accurately with predictions at the temperatures concerned,
though some measure of agreement seems apparent.

\bigskip

We acknowledge valuable discussions with C.P. Korthals Altes, M. Teper
and A. Michels. This work was supported by PPARC grants GR/J21354 and
GR/K55752, and S.T.W. acknowledges the award of a PPARC studentship.

\bigskip
\immediate\closeout\rfile\writestoppt
{{\noindent\bf References}}
\bigskip{\frenchspacing
\parindent=20pt\escapechar=` \input
refs.tmp\vfill\eject}\nonfrenchspacing
\end